\documentclass[12pt,onecolumn]{IEEEtran}
\usepackage{amsmath}
\usepackage{amssymb}
\usepackage{amsfonts}
\usepackage{graphicx}

\input{tcilatex}
\begin{document}

\begin{center}
{\LARGE Output tracking based on extended observer for nonlinear uncertain
systems}

Xinhua Wang, Zengqiang Chen, and Zhuzhi Yuan

{\small Department of Automation, Nankai University, Tianjin, China
(wangxinhua04@gmail.com)}
\end{center}

\emph{Abstract:} A high-gain extended observer is designed for a class of
nonlinear uncertain systems. This observer has the ability of estimating
system uncertainty, and it can be used to estimate the derivatives of signal
up to order $n$. The controller based on this extended observer can make the
tracking error and its derivatives converge to zero rapidly even when
uncertainties and disturbances exist. The result of simulation indicates
that this method has satisfactory control performance for nonlinear
uncertain systems.

\emph{Keywords:} nonlinear uncertain system, high-gain extended observer,
estimating uncertainty, tracking error

\bigskip

\markboth{Control and Decision, 19(10), 2004, pp. 1113-1116}I\textbf{1.
Introduction}

Nonlinear control is a main research field in control theory and engineering
[1-5]. Output tracking problem of nonlinear uncertain systems is a hot topic
of current research, and many control methods have been proposed. In
practice, the derivatives of the tracked signal are generally unknown, which
increases the difficulty of controller design. Output tracking problems with
the assumption of known tracked signal and its derivatives were considered
in [3-5]. Many state observers are poor at observing nonlinear systems and
tend to converge slowly. In [3], a high-gain observer and a sliding mode
control are used for output feedback of nonlinear systems, and the
derivatives of tracked signal are assumed to be known. However, for this
method, the obvious tracking error exists for tracking control of uncertain
systems. In [6,7], the proposed extended state observer has precise
estimation performance and strong ability of disturbance rejection. However,
the system stability was not considered.

In this paper, for a class of nonlinear uncertain systems, a high-gain
extended observer is presented to estimate the system uncertainty and the
unknown states. The observer has rapid convergence rate and accurate
estimation, and the system stability of the extended observer is proved.
Furthermore, a controller is designed based on the extended observer to make
the convergence rate and accuracy of output tracking errors meet the control
requirements.

\bigskip

\textbf{2. Problem analysis}

The following nonlinear uncertain system is considered:

\begin{equation}
\left\{ 
\begin{array}{c}
x^{\left( n\right) }=f\left( x,\dot{x},\cdots ,x^{\left( n-1\right)
},t\right) +b\cdot u, \\ 
\multicolumn{1}{l}{y=x}%
\end{array}%
\right.
\end{equation}%
where, the function $f\left( x,\dot{x},\cdots ,x^{\left( n-1\right)
},t\right) $ includes the uncertainties and disturbances, and its
first-order derivative exists; $u$ is the control input; $b$ is a non-zero
constant; the reference signal is $y_{d}$, and its derivatives are unknown.

Define $x_{1}=x$, $x_{2}=\dot{x}$, $x_{3}=\dot{x}_{2}=x^{\left( 2\right) }$,
and $x_{n}=\dot{x}_{n-1}=x^{\left( n-1\right) }$. Then, $\dot{x}%
_{n}=x^{\left( n\right) }=f\left( x_{1},x_{2},\cdots ,x_{n},t\right) +b\cdot
u$.

Also, define $x_{n+1}=f\left( x_{1},x_{2},\cdots ,x_{n},t\right) $, $\dot{x}%
_{n+1}=f^{\left( 1\right) }\left( x_{1},x_{2},\cdots ,x_{n},t\right)
=g\left( x_{1},x_{2},\cdots ,x_{n},t\right) $, and $\left\vert g\left(
x_{1},x_{2},\cdots ,x_{n},t\right) \right\vert +\left\vert y_{d}^{\left(
n+1\right) }\right\vert \leq M$, where, $0\leq M<+\infty $. Therefore, the
system (1) can be expressed by

\begin{equation}
\left\{ 
\begin{array}{l}
\dot{x}_{1}=x_{2} \\ 
\vdots \\ 
\dot{x}_{n-1}=x_{n} \\ 
\dot{x}_{n}=x_{n+1}+b\cdot u \\ 
\dot{x}_{n+1}=g\left( x_{1},x_{2},\cdots ,x_{n},t\right) \\ 
y=x_{1}%
\end{array}%
\right.
\end{equation}%
Then, the system error is

\begin{equation}
\left\{ 
\begin{array}{l}
\dot{e}_{1}=e_{2} \\ 
\vdots \\ 
\dot{e}_{n-1}=e_{n} \\ 
\dot{e}_{n}=e_{n+1}+b\cdot u \\ 
\dot{e}_{n+1}=g\left( x_{1},x_{2},\cdots ,x_{n},t\right) -y_{d}^{\left(
n+1\right) }%
\end{array}%
\right.
\end{equation}%
where,

\begin{eqnarray*}
e\left( t\right) &=&\left[ 
\begin{array}{ccc}
e_{1} & \cdots & e_{n+1}%
\end{array}%
\right] ^{T} \\
&=&\left[ 
\begin{array}{cccc}
x_{1}-y_{d} & \cdots & x_{n}-y_{d}^{\left( n-1\right) } & 
x_{n+1}-y_{d}^{\left( n\right) }%
\end{array}%
\right] ^{T}
\end{eqnarray*}%
In (2), subtracting $y_{d}^{\left( n\right) }$ from both sides, we can get $%
\dot{e}_{n}=e_{n+1}+b\cdot u$ in (3). $e_{1}=x_{1}-y_{d}$ is known, and the
other variables are unknown.

\bigskip

\textbf{3. Design of extended observer}

Choose the positive real numbers $\lambda _{1}$, $\lambda _{2}$, $\cdots $, $%
\lambda _{n+1}$ that are not equal to each other, and make $\underset{i=1}{%
\overset{n+1}{\Pi }}\left( s+\lambda _{i}\right) =0$ and $%
s^{n+1}+h_{1}s^{n}+\cdots +h_{n}s+h_{n+1}=0$ equal. Define

\begin{eqnarray}
\lambda &=&\min \left\{ \lambda _{1},\lambda _{2},\cdots ,\lambda
_{n+1}\right\} ,  \notag \\
A &=&\left[ 
\begin{array}{cccc}
-\frac{h_{1}}{\varepsilon } & 1 & \cdots & 0 \\ 
\vdots &  & \ddots &  \\ 
-\frac{h_{n}}{\varepsilon ^{n}} & 0 & \cdots & 1 \\ 
-\frac{h_{n+1}}{\varepsilon ^{n+1}} & 0 & \cdots & 0%
\end{array}%
\right] ,\text{ }B=\left[ 
\begin{array}{c}
0 \\ 
\vdots \\ 
0 \\ 
1%
\end{array}%
\right] \\
\varepsilon &\in &\left( 0,1\right)  \notag
\end{eqnarray}

For the above matrix $A$, there exist the Vandermonde matrix $T$ such that

\begin{equation}
A=T\cdot \text{diag}\left[ -\frac{h_{1}}{\varepsilon },\cdots ,-\frac{h_{n+1}%
}{\varepsilon ^{n+1}}\right] \cdot T^{-1}
\end{equation}

\bigskip

\textbf{Theorem 1:} For the system error (3), the extended observer is
designed as follows:

\begin{equation}
\left\{ 
\begin{array}{l}
\dot{\widehat{e}}_{1}=\widehat{e}_{2}-\frac{h_{1}}{\varepsilon }\left( 
\widehat{e}_{1}-e_{1}\right) \\ 
\multicolumn{1}{c}{\vdots} \\ 
\dot{\widehat{e}}_{n}=\widehat{e}_{n+1}-\frac{h_{n}}{\varepsilon ^{n}}\left( 
\widehat{e}_{1}-e_{1}\right) +b\cdot u \\ 
\dot{\widehat{e}}_{n+1}=-\frac{h_{n+1}}{\varepsilon ^{n+1}}\left( \widehat{e}%
_{1}-e_{1}\right)%
\end{array}%
\right.
\end{equation}

Therefore, we get the following conclusions:

1)

\begin{equation}
\underset{\varepsilon \rightarrow 0^{+}}{\lim }\left\Vert \delta \left(
t\right) \right\Vert =0
\end{equation}

2) When a $\varepsilon \in \left( 0,1\right) $ is selected, we get

\begin{equation}
\underset{t\rightarrow +\infty }{\lim }\left\Vert \delta \left( t\right)
\right\Vert \leq \frac{\varepsilon M}{\lambda }\left\Vert T\right\Vert
\left\Vert T^{-1}\right\Vert
\end{equation}%
where,

\begin{eqnarray*}
\delta \left( t\right) &=&\left[ 
\begin{array}{ccc}
\delta _{1} & \cdots & \delta _{n+1}%
\end{array}%
\right] ^{T} \\
&=&\left[ 
\begin{array}{ccc}
\widehat{e}_{1}-e_{1} & \cdots & \widehat{e}_{n+1}-e_{n+1}%
\end{array}%
\right] ^{T}, \\
\widehat{e}\left( t\right) &=&\left[ 
\begin{array}{ccc}
\widehat{e}_{1} & \cdots & \widehat{e}_{n+1}%
\end{array}%
\right] ^{T},
\end{eqnarray*}%
$\varepsilon \in \left( 0,1\right) $ is the perturbation parameter.

\emph{Proof:} The system error between (4) and (3) is

\begin{equation}
\dot{\delta}\left( t\right) =A\delta \left( t\right) +B\left( -g\left(
x_{1},x_{2},\cdots ,x_{n},t\right) -y_{d}^{\left( n+1\right) }\right)
\end{equation}%
Then, the solution to (9) can be expressed by

\begin{equation}
\delta \left( t\right) =\exp \left( A\cdot t\right) \delta \left( 0\right)
+\int_{0}^{t}\exp \left( A\left( t-\tau \right) \right) \left( -g\left(
x_{1},x_{2},\cdots ,x_{n},t\right) -y_{d}^{\left( n+1\right) }\right) d\tau B
\end{equation}%
Therefore, we get

\begin{eqnarray}
\delta \left( t\right) &=&\left\Vert \exp \left( A\cdot t\right) \right\Vert
\left\Vert \delta \left( 0\right) \right\Vert +M\left\Vert \int_{0}^{t}\exp
\left( A\left( t-\tau \right) \right) d\tau \right\Vert \left\Vert
B\right\Vert  \notag \\
&\leq &\left\Vert T\cdot \text{diag}\left\{ \exp \left( -\frac{\lambda _{1}}{%
\varepsilon }t\right) ,\cdots ,\exp \left( -\frac{\lambda _{n+1}}{%
\varepsilon }t\right) \right\} \cdot T^{-1}\right\Vert \left\Vert \delta
\left( 0\right) \right\Vert  \notag \\
&&+M\int_{0}^{t}\left\Vert \exp \left( A\left( t-\tau \right) \right)
\right\Vert d\tau \left\Vert B\right\Vert  \notag \\
&\leq &\left\Vert T\right\Vert \left\Vert T^{-1}\right\Vert \exp \left( -%
\frac{\lambda }{\varepsilon }t\right) \left\Vert \delta \left( 0\right)
\right\Vert  \notag \\
&&+\left\Vert T\right\Vert \left\Vert T^{-1}\right\Vert M\int_{0}^{t}\exp
\left( -\frac{\lambda }{\varepsilon }\left( t-\tau \right) \right) d\tau
\left\Vert B\right\Vert  \notag \\
&\leq &\left\Vert T\right\Vert \left\Vert T^{-1}\right\Vert \left[ \exp
\left( -\frac{\lambda }{\varepsilon }t\right) \left\Vert \delta \left(
0\right) \right\Vert +M\frac{\varepsilon }{\lambda }\left( 1-\exp \left( 
\frac{\lambda }{\varepsilon }t\right) \right) \left\Vert B\right\Vert \right]
\end{eqnarray}%
Because $\left\Vert B\right\Vert =1$ and $\left\Vert \delta \left( 0\right)
\right\Vert $ is bounded, $\underset{\varepsilon \rightarrow 0^{+}}{\lim }%
\left\Vert \delta \left( t\right) \right\Vert =0$. When $\varepsilon \in
\left( 0,1\right) $ is selected, from (11), we can get $\underset{%
t\rightarrow +\infty }{\lim }\left\Vert \delta \left( t\right) \right\Vert
\leq \frac{\varepsilon M}{\lambda }\left\Vert T\right\Vert \left\Vert
T^{-1}\right\Vert $. This concludes the proof. $\blacksquare $

The aim of observer design is to make $\widehat{e}_{1}\rightarrow e_{1}$, $%
\cdots $, $\widehat{e}_{n+1}\rightarrow e_{n+1}$. This extended observer can
be used to estimate the system uncertainty and signal derivatives up to
order $n$. Although the convergence speed of common linear extended
observers are slower than that of nonlinear extended observers in the
neighborhood of the equilibrium, the use of high gains in the observer can
speed up the convergence. In addition to estimation of unknown error
variables and system uncertainty in (3), the controller $u$ is designed
according to the observer estimation to implement the system tracking.

\bigskip

\textbf{4. Controller design}

\textbf{Theorem 2:} For the system error (3) and the extended observer (6),
a sliding variable is select as

\begin{equation}
\sigma \left( t\right) =\widehat{e}_{n}+a_{n-1}\widehat{e}_{n-1}+\cdots
+a_{1}\widehat{e}_{1}
\end{equation}%
where, the polynomial $s^{n-1}+a_{n-1}s^{n-2}+\cdots +a_{1}=0$ is Hurwitz.
The controller is designed as

\begin{eqnarray}
u &=&-b^{-1}\left( U_{0}\text{sign}\left( \sigma \left( t\right) \right)
-\left( \frac{h_{n}}{\varepsilon ^{n}}+a_{n-1}\frac{h_{n-1}}{\varepsilon
^{n-1}}+\cdots +a_{1}\frac{h_{1}}{\varepsilon ^{1}}\right) \left( \widehat{e}%
_{1}\rightarrow e_{1}\right) \right.  \notag \\
&&\left. +\widehat{e}_{n+1}+a_{n-1}\widehat{e}_{n}+\cdots +a_{1}\widehat{e}%
_{2}\right)
\end{eqnarray}%
Then, we can get

\begin{equation}
\underset{t\rightarrow \infty }{\lim }\left\Vert e\left( t\right)
\right\Vert \leq k_{p}\sqrt{\varepsilon }
\end{equation}%
where, $k_{p}$ and $U_{0}$ are the positive constants.

\emph{Proof:} Select a Lyapunov function candidate as $V=\frac{1}{2}\sigma
^{2}\left( t\right) $. Then, we get

\begin{eqnarray}
\dot{V} &=&\sigma \left( t\right) \left( \dot{\widehat{e}}_{n}+a_{n-1}\dot{%
\widehat{e}}_{n-1}+\cdots +a_{1}\dot{\widehat{e}}_{1}\right)  \notag \\
&=&\sigma \left( t\right) \left( \widehat{e}_{n+1}-\frac{h_{n}}{\varepsilon
^{n}}\left( \widehat{e}_{1}-e_{1}\right) +bu+a_{n-1}\left( \widehat{e}_{n}-%
\frac{h_{n-1}}{\varepsilon ^{n-1}}\left( \widehat{e}_{1}-e_{1}\right)
\right) +\cdots +a_{1}\left( \widehat{e}_{2}-\frac{h_{1}}{\varepsilon }%
\left( \widehat{e}_{1}-e_{1}\right) \right) \right)  \notag \\
&=&-U_{0}\left\vert \sigma \left( t\right) \right\vert =-\sqrt{2}U_{0}V^{%
\frac{1}{2}}
\end{eqnarray}%
Therefore, there exist a finite time $T_{0}$, for $t\geq T_{0}$, the
variables are in the sliding surface $\sigma \left( t\right) =0$ [8]. From
(3) and $\sigma \left( t\right) =0$, for $t\geq T_{0}$, we get

\begin{eqnarray}
\dot{e}_{n-1} &=&e_{n}=\widehat{e}_{n}-\delta _{n}=-\left( a_{n-1}\widehat{e}%
_{n-1}+\cdots +a_{1}\widehat{e}_{1}\right) -\delta _{n}  \notag \\
&=&-\left\{ a_{1}\left( e_{1}+\delta _{1}\right) +\cdots +a_{n-1}\left(
e_{n-1}+\delta _{n-1}\right) \right\} -\delta _{n}  \notag \\
&=&-a_{1}e_{1}-\cdots -a_{n-1}e_{n-1}-a_{1}\delta _{1}-\cdots -a_{n-1}\delta
_{n-1}-\delta _{n}
\end{eqnarray}%
From (3) and (16), we get

\begin{equation}
\dot{\widetilde{e}}=\widetilde{A}\cdot \widetilde{e}+H\cdot \delta \left(
t\right)
\end{equation}%
where,

\begin{eqnarray}
\widetilde{e} &=&\left[ 
\begin{array}{ccc}
e_{1} & \cdots & e_{n-1}%
\end{array}%
\right] ^{T},\text{ }\delta \left( t\right) =\left[ 
\begin{array}{ccc}
\delta _{1} & \cdots & \delta _{n+1}%
\end{array}%
\right] ^{T},  \notag \\
\widetilde{A} &=&\left[ 
\begin{array}{cccc}
0 & 1 & \cdots & 0 \\ 
\vdots & \vdots & \ddots & \vdots \\ 
0 & \cdots &  & 1 \\ 
-a_{1} & -a_{2} & \cdots & -a_{n-1}%
\end{array}%
\right] ,  \notag \\
H &=&\left[ 
\begin{array}{ccccc}
0 & \cdots & 0 & 0 & 0 \\ 
\vdots & \ddots & \vdots & \vdots & \vdots \\ 
0 & \cdots & 0 & 0 & 0 \\ 
-a_{1} & \cdots & -a_{n-1} & -1 & 0%
\end{array}%
\right]
\end{eqnarray}%
Because both $\widetilde{A}$ and $A$ are Hurwitz, for the given
positive-definite matrices $Q_{1}$ and $Q_{2}$, there exist the
positive-define matrices $P_{1}$ and $P_{2}$, such that

\begin{equation}
P_{1}\widetilde{A}+\widetilde{A}^{T}P_{1}=-Q_{1}\text{, }%
P_{2}A+A^{T}P_{2}=-Q_{2}
\end{equation}

Define $\Phi \left( \widetilde{e},\text{ }\delta \left( t\right) \right) =%
\widetilde{e}^{T}P_{1}\widetilde{e}+\delta ^{T}\left( t\right) P_{2}\delta
\left( t\right) $. Taking derivative for $\Phi \left( \widetilde{e},\text{ }%
\delta \left( t\right) \right) $ along the solutions of equations (9) and
(17), we get

\begin{equation}
\dot{\Phi}\left( \widetilde{e},\text{ }\delta \left( t\right) \right) <-\eta
_{1}\Phi \left( \widetilde{e},\text{ }\delta \left( t\right) \right) ,\text{ 
}\Phi \left( \widetilde{e},\text{ }\delta \left( t\right) \right)
>r_{1}\varepsilon
\end{equation}%
where, $\eta _{1}$ and $r_{1}$ are the positive constants. Select $%
r_{2}>r_{1}$, and define

\begin{equation}
\Omega =\left\{ \left. \widetilde{e},\text{ }\delta \left( t\right)
\right\vert \Phi \left( \widetilde{e},\text{ }\delta \left( t\right) \right)
\leq r_{2}\varepsilon \right\}
\end{equation}%
Then, we can find that, there exists a finite time $t_{1}$, for $t>t_{1}$,\
such that $\Phi \left( \widetilde{e},\text{ }\delta \left( t\right) \right)
\in \Omega $. Therefore, from (16) and $\underset{t\rightarrow +\infty }{%
\lim }\left\Vert \delta \left( t\right) \right\Vert \leq \frac{\varepsilon M%
}{\lambda }\left\Vert T\right\Vert \left\Vert T^{-1}\right\Vert $, we can
get $\underset{t\rightarrow \infty }{\lim }\left\Vert e\left( t\right)
\right\Vert \leq k_{p}\sqrt{\varepsilon }$, where, $k_{p}$ is a positive
constant. This concludes the proof. $\blacksquare $

\bigskip

\textbf{5. Simulation example}

The following system is considered:

\begin{eqnarray*}
\dot{x}_{1} &=&x_{2} \\
\dot{x}_{2} &=&\cos \frac{\pi }{2}x_{1}-x_{1}^{1/3}-4x_{2}^{1/3}+u \\
y &=&x_{1}
\end{eqnarray*}

The reference signal $y_{d}=2\sin t$. From (2), (3), (6) and $e_{1}=y-y_{d}$%
, the designed extended observer is as follows:

\begin{equation*}
\begin{array}{l}
\dot{\widehat{e}}_{1}=\widehat{e}_{2}-\frac{6}{0.1}\left( \widehat{e}%
_{1}-e_{1}\right) \\ 
\dot{\widehat{e}}_{2}=\widehat{e}_{3}-\frac{11}{0.1^{2}}\left( \widehat{e}%
_{1}-e_{1}\right) +u \\ 
\dot{\widehat{e}}_{3}=-\frac{6}{0.1^{3}}\left( \widehat{e}_{1}-e_{1}\right)%
\end{array}%
\end{equation*}

Select the sliding variable $\sigma \left( t\right) =\widehat{e}_{2}+%
\widehat{e}_{1}$, and the controller is designed as

\begin{equation*}
u=-\left( 4\text{sign}\left( \sigma \left( t\right) \right) -\left( \frac{11%
}{0.1^{2}}+\frac{6}{0.1}\right) \left( \widehat{e}_{1}-e_{1}\right) +%
\widehat{e}_{3}+\widehat{e}_{2}\right)
\end{equation*}

The plot of output tracking errors is shown in Figure 1.

\begin{figure}[tbp]
\centering
\includegraphics[width=3.00in]{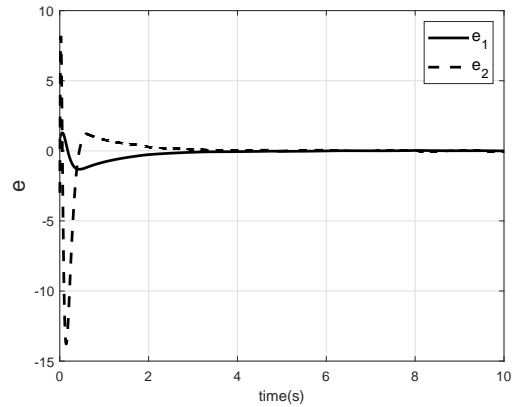}
\caption{System output tracking errors.}
\end{figure}

\bigskip

\textbf{6. Conclusion}

This paper has presented a method of output tracking control based on
extended observer. From the theoretical analysis and simulation, the
convergence rate and precision of estimation and control are satisfactory.
The future job is to design observer and controller for nonlinear
non-minimum-phase systems using extended observer and the centre-manifold
theory.

\end{document}